\documentclass[twocolumn]{revtex4}
\usepackage{}
\usepackage{epsfig,latexsym,amssymb}
\usepackage{latexsym}
\usepackage{amsmath}
\usepackage{amssymb}
\usepackage{amsfonts}
\usepackage{comment}
\usepackage{graphicx}
\usepackage{verbatim}
\usepackage{epstopdf}
\usepackage{hyperref, url}
\hypersetup{colorlinks   = true,
            urlcolor     = blue,
            citecolor    = blue,
            linkcolor    = blue,
            menucolor    = blue,
            anchorcolor  = blue,
            filecolor    = blue}

\newcommand{\be}{\begin{equation}}
\newcommand{\ee}{\end{equation}}
\newcommand{\bea}{\begin{eqnarray}}
\newcommand{\eea}{\end{eqnarray}}

\begin{document}

\title{Can Planet 9 be an Axion Star?}

\author{Haoran Di}
\email{hrdi@ecut.edu.cn}
\affiliation{School of Science, East China University of Technology, Nanchang 330013, China}

\author{Haihao Shi}
\affiliation{School of Science, East China University of Technology, Nanchang 330013, China}

\begin{abstract}
The anomalous orbits of Trans-Neptunian Objects (TNOs) can be explained by the Planet 9 hypothesis. We propose that the Planet 9 can be an axion star. Axion stars are gravitational bound clusters condensed by QCD axions or axion-like particles (ALPs), which we call axions for brevity. We find that the probability of capturing an axion star by the solar system is the same order of magnitude as the probability of capturing a free floating planet (FFP), and even higher for the case of axion star, with axion star mass $5M_\oplus\approx1.5\times10^{-5}M_\odot$ and $\Omega_{\rm{AS}}/\Omega_{\rm{DM}}\simeq 1/10$. Although axion star can emit monochromatic signals through two-photon decay, we find that the frequency of decay photon is either not within the frequency range of the radio telescope, or the decay signal is too weak to be detected. Therefore, if Planet 9 is composed by an axion star, it may be difficult to distinguish it from an isolated primordial black hole by spontaneous decay of axion.
\end{abstract}

\maketitle

\section{Introduction}
Mounting evidence from various observations and theoretical models has led to the realization that dark matter constitutes a substantial portion of the total energy density. However, the nature and composition of dark matter particles remain shrouded in mystery. The discovery and confirmation of the Higgs boson's existence \cite{CMS:2012qbp,ATLAS:2012yve} have sparked a resurgence of interest in boson stars. The existence of other scalar fields stemming from theories beyond the Standard Model of particle physics has further fueled curiosity. For instance, the Peccei-Quinn mechanism \cite{Peccei:1977ur,Peccei:1977hh}, a prominent solution to the strong-CP problem in quantum chromodynamics (QCD), predicts the existence of the QCD axion \cite{Weinberg:1977ma,Wilczek:1977pj}. String theory provides compelling motivations for the existence of a vast multitude of axion-like particles (ALPs) spanning a wide range of mass scales, giving rise to what is known as the ``axiverse" \cite{Arvanitaki:2009fg}.  ALPs can also arise from ``$\pi$-axiverse'' \cite{Alexander:2023wgk}. For brevity, we will refer to QCD axions and ALPs as axions. The axions can be produced by misalignment mechanism \cite{Preskill:1982cy,Abbott:1982af,Dine:1982ah},
string defects \cite{Gorghetto:2020qws}, or kinetic misalignment mechanism \cite{Co:2019jts}, etc. Given their bosonic nature, axions possess the ability to attain remarkably high phase space density, leading to the intriguing phenomenon of Bose-Einstein condensation (BEC) \cite{Sikivie:2009qn}. As a consequence of BEC, axions can aggregate into gravitationally bound structures known as axion stars. See Refs. \cite{Braaten:2019knj,Visinelli:2021uve} for recent reviews. Importantly, it is plausible that a considerable portion of the elusive axion dark matter \cite{Preskill:1982cy,Abbott:1982af,Dine:1982ah} manifests itself in the form of these captivating axion stars.

The anomaly orbits of trans-Neptunian objects (TNOs) \cite{Brown:2004yy,Trujillo,Batygin:2016zsa} and gravitational anomalies observed
by the Optical Gravitational Lensing Experiment (OGLE) \cite{Mroz} need to be explained. The Planet 9, with mass $M_9\sim5-15M_\oplus$ and distance $300-1000\rm{AU}$ from the Sun, was proposed to explain the anomalies connected to the orbits of TNOs \cite{Batygin-PR}.  The free floating planets (FFPs) or
primordial black holes (PBHs) with mass $M\sim0.5-20M_\oplus$ \cite{Niikura:2019kqi} can explain the gravitational anomalies observed by the OGLE. It is an interesting possibility that Planet 9 can be explained by the PBH captured by solar system \cite{Scholtz:2019csj} or planetary mass black hole due to heavy dark matter accretion into planet \cite{Ray:2023auh}.
We consider another interesting possibility: Planet 9 is composed of axion star and captured by the solar system. In this paper we use the natural units, $c = \hbar=1$.

\section{Axion Stars}
Although a significant amount of research has been conducted on compact objects made up of fermions, such as neutron stars and white dwarfs, there is also an ongoing endeavor to investigate localized solitons composed of bosons. This quest can be traced back to Wheeler's initial concept of electromagnetic ``geons" \cite{Wheeler:1955zz}. Complex scalar fields possess the capacity to aggregate and lead to the formation of compact entities recognized as boson stars \cite{Kaup:1968zz}. Contrasting a real scalar field, which lacks a U(1) symmetry essential for the presence of a conserved charge, stability becomes less certain. Nevertheless, even in the absence of a clearly conserved Noether current, solutions in the form of solitons for the scalar boson field equation persist, referred to as oscillatons \cite{Seidel:1991zh,Copeland:1995fq}.

The QCD axion \cite{Weinberg:1977ma,Wilczek:1977pj}, characterized as a spin-0 pseudoscalar boson, possesses a minute mass labeled as $m_\phi$. It exhibits exceedingly feeble self-interaction, and extremely weak interaction with Standard Model particles.
The need for the Lagrangian to maintain shift symmetry invariance results in the axion potential $V(\phi)$ displaying cyclic behavior in relation to $\phi$:
\bea
V(\phi)=V(\phi+2\pi f_a),
\eea
where $f_a$ is called axion decay constant representing the energy scale of spontaneous breaking of the $U(1)$ symmetry. The simplest model employed in the majority of axion's phenomenological investigations is the instanton potential \cite{Peccei:1977ur}:
\bea
V(\phi)=(m_\phi f_a)^2[1-\cos(\phi/f_a)].
\eea
The leading self-interaction term in the expansion of $V(\phi)$ is $\lambda\phi^4/4!$ with attractive coupling $\lambda=-m_\phi^2/f_a^2$.
Due to their bosonic nature, axions have the potential to achieve remarkably high phase space density, leading to the formation of a BEC \cite{Sikivie:2009qn}. This condensate can give rise to gravitationally bound configurations called axion stars, which can be dilute or dense \cite{Chavanis:2017loo,Visinelli:2017ooc,Eby:2019ntd}. Gravitational cooling plays a significant role in facilitating the relaxation of axion stars towards a stable configuration \cite{Seidel:1993zk,Guzman:2006yc}. While axion stars are generally self-gravitating objects \cite{Barranco:2010ib}, their equilibrium state is significantly influenced by self-interactions.  Axion stars reach a stable state through a balance of kinetic pressure, gravitational attraction, and self-interactions \cite{Eby:2016cnq,Kaup:1968zz,Ruffini:1969qy,Breit:1983nr,Colpi:1986ye,Seidel:1990jh,Friedberg:1986tq,Seidel:1991zh,Liddle:1992fmk,Chavanis:2011zi,Chavanis:2011zm}.

Within dilute axion stars, self-interactions can generally be overshadowed by gravity, while in more compact stars, self-interactions emerge as the predominant influence over the axion star's dynamics. If a dilute axion star attains additional axions and subsequently collapses beyond its critical mass, it has the potential to trigger a phenomenon called ``bosenova" \cite{Eby:2016cnq,Levkov:2016rkk}, ultimately resulting in a denser axion star as the residual entity, with a mass ranging from $10^{-20}M_\odot$ to about $M_\odot$ \cite{Braaten:2015eeu}. Nevertheless, the temporal span of a dense axion star's existence might be too brief to grant it cosmological significance as an astrophysical object \cite{Braaten:2019knj,Visinelli:2017ooc,Hertzberg:2010yz,Eby:2015hyx}.

In the conventional post-inflationary scenario, the spontaneous breaking of the U(1) symmetry takes place subsequent to the inflationary period. The fraction of dark matter occupied by axion stars can be as high as $75\%$ for QCD axions \cite{Eggemeier:2019khm,Eggemeier:2022hqa}, but for other models, this fraction is much less known. In addition, observations of microlensing events within the HSC and OGLE data consistently point to about $27^{+7}_{-13}$ percent of dark matter potentially existing in the form of axion stars \cite{Sugiyama:2021xqg}. Therefore, a segment of dark matter might exist in the configuration of axion stars, which may affect the current detection experiments of axions. Because of the existing uncertainties, we assume that the proportion of axion stars to dark matter is approximately 1/10 in the following discussion.

\section{Capture Probability of Axion Stars}
There are three potential explanations for the origin of Planet 9: (i) Planet 9 originated in its current orbit; (ii) Planet 9 formed closer to the Sun and was subsequently scattered into its current orbit; or (iii) Planet 9 formed outside of the Solar System and was later captured. Although the chances of all three scenarios happening are slim, there is a comparable level of probability between capturing an Earth-mass PBH and capturing an FFP of similar mass \cite{Scholtz:2019csj}. The capture probability of the solar system can be represented in the following way:
\bea
\Gamma=n_{\rm{AS}} \langle\sigma  v\rangle= n_{\rm{AS}} \int  f(v+v_{\odot,{\rm{AS}}})\frac{d\sigma}{dv}v dv,
\eea
where $n_{\rm{AS}}$ and $v$ represent the number density and velocity of axion star, $d\sigma/dv$ is the differential capture cross section, $f(v)$ is velocity distribution, and $v_{\odot,{\rm{AS}}}$ is the average velocity of the Sun towards axion star.

The differential capture cross section is identical for FFPs, PBHs or axion stars and greatly lowered when the relative velocity is larger than $0.4{\rm{km/s}}$ \cite{Goulinski}, which is much smaller than the relative velocity of the Sun towards axion star. Therefore, the velocity distribution of axion star can be approximated by the zero order value $f(v_{\odot,{\rm{AS}}})$. Then the ratio of capturing axion star to FFP approximate as follows
\bea \label{ratio}
\frac{\Gamma_{\rm{AS}}}{\Gamma_{\rm{FFP}}}\simeq \frac{n_{\rm{AS}}}{n_{\rm{FFP}}}
\frac{f_{\rm{AS}}(v_{\odot,{\rm{AS}}})}{f_{\rm{FFP}}(v_{\odot,{\rm{FFP}}})}.
\eea
We assume $v_{\odot,{\rm{AS}}}$  and velocity dispersion $\sigma_{\odot,{\rm{AS}}}$ are same as the relative velocity and velocity dispersion of solar system to dark matter halo (DMH), with $v_{\rm{DMH}}=220{\rm{km/s}}$ and $\sigma_{\rm{DMH}}=v_{\rm{DMH}}/\sqrt2$ \cite{Drukier:1986tm}.
Considering FFP captured in the field far away from the star formation region, the number density of FFP $n_{\rm{FFP}}$ is about $0.24{\rm{pc}^{-3}}$  from microlensing surveys \cite{Thies:2011cb}. The local density of dark matter near the solar system is $\rho=0.38\rm{GeV/cm^3}$ \cite{McMillan:2016}. Therefore, we can obtain the local density of axion stars:
\bea \label{density}
n_{\rm{AS}}=\frac{\Omega_{\rm{AS}}}{\Omega_{\rm{DM}}}\frac{\rho_{\rm{DM}}}{M_{\rm{AS}}}
\sim 70{\rm{pc}}^{-3}\left(\frac{10 \Omega_{\rm{AS}}}{\Omega_{\rm{DM}}}\right)
\left(\frac{5 M_\oplus}{M_{\rm{AS}}}\right).
\eea
The velocity distribution is modeled as a Gaussian distribution
\bea \label{distribution}
f(v)=\frac{1}{\left(2\pi\sigma^2\right)^{3/2}}e^{-v^2/{2\sigma^2}},
\eea
where $\sigma$ is the velocity dispersion.
Substituting Eq. \eqref{density} and Eq. \eqref{distribution} into Eq. \eqref{ratio}, the ratio of axion star to FFP capture probability change to
\bea
\frac{\Gamma_{\rm{AS}}}{\Gamma_{\rm{FFP}}}&\sim& 4.97
\left(\frac{0.24 {\rm{pc}}^{-3}}{n_{\rm{FFP}}}\right)
\left(\frac{10 \Omega_{\rm{AS}}}{\Omega_{\rm{DM}}}\right)\nonumber\\
&&\times\left(\frac{5 M_\oplus}{M_{\rm{AS}}}\right)
\left(\frac{\sigma_{\rm{FFP}}}{40{\rm{km/s}}}\right)^3,
\eea
where the velocity dispersion of FFP is assumed to be about $40{\rm{km/s}}$ in the thin disc which can be considered as a plausible source of FFP \cite{Goulinski}.
From the above equation, it can be seen that the probability of capturing an axion star is the same order of magnitude as the probability of capturing an FFP, and even higher for the case of axion star, with axion star mass $5M_\oplus\approx1.5\times10^{-5}M_\odot$ and $\Omega_{\rm{AS}}/\Omega_{\rm{DM}}\simeq 1/10$. Therefore, Planet 9 could be an axion star captured by the solar system.

\section{Quantum Decay from Dilute Axion Stars}

Ignoring the self-interaction term for dilute axion star, the general Lagrangian of QCD axion and ALP can be written as
\bea \label{axion}
{\cal L}&=&{1\over 2}\partial_{\mu}\phi\partial^{\mu}\phi-{m_{\phi}^2\over 2}\phi^2
+{1 \over4}g_{a\gamma\gamma}\phi F_{\mu\nu} \tilde F^{\mu\nu},
\eea
where $F_{\mu\nu}$ represents the electromagnetic field tensor, $\tilde F^{\mu\nu}=\frac{1}{2} F_{\alpha\beta} \epsilon^{\mu\nu\alpha\beta}$ is the dual tensor of $F_{\mu\nu}$, $\phi$ is the axions field, and the axion-photon coupling $g_{a\gamma\gamma}=\alpha g/(2\pi f_a)$  is related to the
axion decay constant $f_a$. $\alpha$ is the fine structure constant, and $g$ is a model-dependent constant of order one, which we will set $g=1$ in the following discussion. For QCD axions, the decay constant $f_a$ is related by the mass of axion: $f_a\simeq6\times10^{11}(10^{-5}{\rm{eV}}/m_{\phi})\rm{GeV}$.

\begin{figure}
\begin{center}
\includegraphics[width=0.5\textwidth]{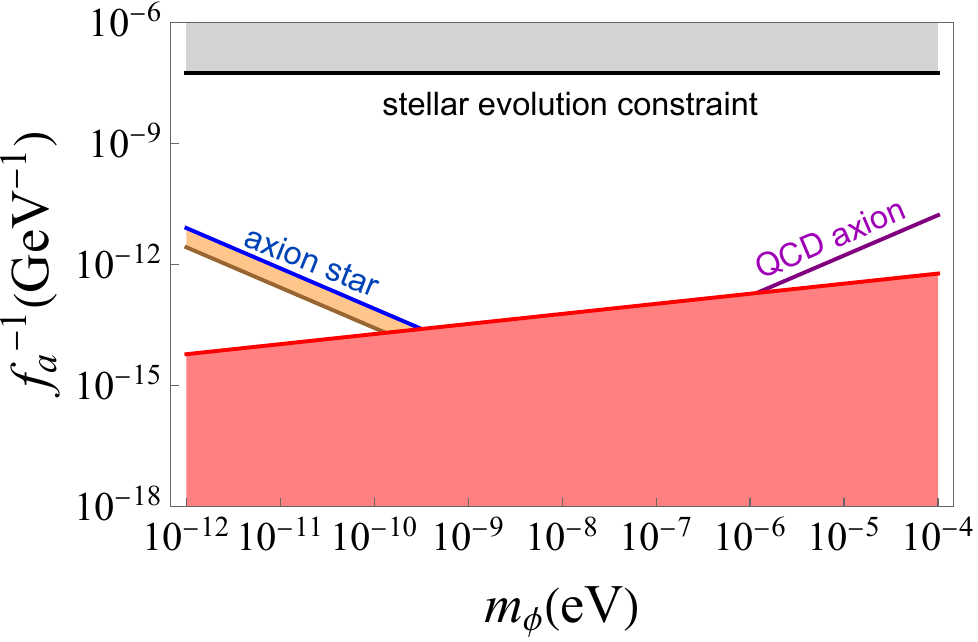}
\caption{Constraints on the parameter space of axions. The solid red line represents the cosmology abundance given by Eq. \eqref{abundance}, and the area below this line is excluded due to the production of too much dark matter. The black solid line is the limit given by stellar evolution \cite{Ayala:2014pea,Giannotti:2015kwo}, and the upper region is excluded. The narrow brown band indicates the possible parameter space of the axion to be the Planet 9 with mass $M_9\sim5-15M_\oplus$. The QCD axion is limited to the purple line due to the connection between its mass and $f_a$. Note that the QCD axion star cannot be the Planet 9 due to its insufficient mass $\simeq5.24\times10^{-10}M_\odot$. The possible axion mass range is from $1.32\times10^{-10}{\rm{eV}}$ to $3.19\times10^{-10}{\rm{eV}}$, which is not within the detection frequency range of the radio telescope, such as FAST or SKA.}
\label{fig:constraints}
\end{center}
\end{figure}

The equation governing the motion of the axion, which approximates the early universe scenario by considering only the quadratic segment of the axion potential, can be expressed as:
\bea
\ddot{\phi}+3H\dot{\phi}+m_\phi^2\phi=0,
\eea
where $H$ is the Hubble parameter.
The relic abundance today is decided by the initial values of axion field $\phi_0\sim f_a$. Oscillations begin when the Hubble parameter $H(t)$ becomes comparable to $m_\phi$, ie. $3H(t_{\rm{osc}})\sim2m_\phi$, corresponding to $t_{\rm{osc}} \simeq 3/(4m_{\phi})$. Then the energy density of axions for a given time is
\bea
\rho_\phi=\frac{1}{2}(\phi^2+m_{\phi}^2 \phi^2).
\eea
Subsequently, at the initiation of oscillations, the number density of axions is represented by
\bea
n_\phi(t_{\rm{osc}})=\frac{1}{2}m_{\phi} f_a^2.
\eea
The current axion relic density can be formulated as follows:
\bea \label{omega}
\Omega_\phi&=&\frac{m_\phi n_\phi(t_{\rm{osc}})}{\rho_c}
\left(
\frac{a(t_{\rm{osc}})}{a_0}
\right)^3\nonumber\\
&=&\frac{m_\phi^2 f_a^2}{2\rho_c}
\left(
\frac{T_0}{T_{\rm{osc}}}
\right)^3,
\eea
where critical density $\rho_c\approx 10.54h^2\rm{GeV/m^3}$ and $T_0\approx2.7\rm{K}$ \cite{ParticleDataGroup:2018ovx}. The axion mass is related to the temperature of oscillation by
\bea \label{mass}
m_\phi\sim\frac{3}{2}H(t_{\rm{osc}})=\frac{3}{2}\sqrt{\frac{\pi^2}{90}}\frac{g_\star^{1/2}(T_{\rm{osc}})}{M_{pl}}T_{\rm{osc}}^2.
\eea
By bringing Eq. \eqref{mass} into Eq. \eqref{omega}, we can obtain
\bea \label{abundance}
\Omega_\phi h^2&\simeq&0.12
\left(\frac{g_\star(T_{\rm{osc}})}{106.75}\right)^{3/4}
\left(\frac{m_\phi}{10^{-6}\rm{eV}}\right)^{1/2}\nonumber\\
&&\times\left(\frac{f_a}{5.32\times 10^{12}\rm{GeV}}\right)^2.
\eea
This equation, which is suitable for axions produced from the misalignment mechanism, provides the lower limit of the parameter space of axions, as shown in Fig.~\ref{fig:constraints}.

Axions are not completely stable due to the interaction term ${\cal L}_{int}=1/4 g_{a\gamma\gamma}\phi F_{\mu\nu} \tilde F^{\mu\nu}$. The decay rate of axions into two photons is as follows:
\bea
\Gamma(\phi\rightarrow\gamma\gamma)&=&\frac{\alpha^2 m_\phi^3}{256 \pi^3 f_a^2}\nonumber\\
&=&1.02\times 10^{-11} {\rm{s}^{-1}} \left( \frac{m_\phi}{\rm{eV}} \right)^3 \left(\frac{\rm{GeV}}{f_a}\right)^2.\nonumber
\\
\eea
This means that the lifetime of the axion is
\bea
\tau_\phi=9.80\times 10^{10} {\rm{s}} \left( \frac{\rm{eV}} {m_\phi} \right)^3 \left(\frac{f_a} {\rm{GeV}}\right)^2.
\eea
Due to the relation $f_a\simeq6\times10^{11}(10^{-5}{\rm{eV}}/m_{\phi})\rm{GeV}$, the lifetime of QCD axion becomes
$\tau_\phi=3.53\times 10^{24} {\rm{s}} \left( {m_\phi}/{\rm{eV}}  \right)^{-5}$.
To be a suitable candidate for dark matter, axions should have lifetime longer than the age of the universe. This provides a limitation on the mass of axion $m_\phi\lesssim \rm{few~eV}$.
Although the axion is very stable, the luminosity $L_\phi$ from decay photons of an axion star may be detectable due to the huge particle number $N_\phi$ in the axion star. The luminosity of the axion star can expressed as:
\bea \label{luminosity}
L_{\phi}&=&N_{\phi} m_{\phi} \Gamma(\phi\rightarrow\gamma\gamma)=M_{\phi} \Gamma(\phi\rightarrow\gamma\gamma)\nonumber\\
&=&4.75\times10^{-33}{M_{\phi}\over M_{\odot}}\left({m_{\phi}\over {10^{-6}\rm eV}}\right)^3\nonumber\\
&&\times\left({10^{12}{\rm GeV}\over f_a}\right)^2L_{\odot},
\eea
where $M_\odot$ and $L_\odot$ denote the solar mass and the solar luminosity respectively, $M_\phi$ represents axion star mass.  The spectral line manifests as an almost monochromatic frequency, approximately $f\simeq m_\phi/({4\pi})$, which stands out as a distinct and recognizable signal feature.

The maximum mass and the corresponding minimum radius of an axion star are given by \cite{Chavanis:2017loo,Visinelli:2017ooc,Chavanis:2011zi,Chavanis:2011zm,Schiappacasse:2017ham}
\bea
M_{\rm{AS}}=1.2\times10^{-6}M_{\odot}\left(\frac{m_\phi}{10^{-8}{\rm{eV}}}\right)^{-1}
\left(\frac{f_a}{10^{14}{\rm{GeV}}}\right),
\label{max}
\eea
\bea
R_{\rm{AS}}=7.8\times10^2{\rm{km}}\left(\frac{m_\phi}{10^{-8}{\rm{eV}}}\right)^{-1}
\left(\frac{f_a}{10^{14}{\rm{GeV}}}\right)^{-1}.
\label{min}
\eea
A large parameter space is potentially ruled out due to the formation of axion stars above critical masses \cite{Fox:2023aat}.
If the dominant constituent of dark matter is composed of QCD axions, then the axion's mass would be approximately $1.17\times10^{-6}{\rm{eV}}$, with a corresponding decay constant of $5.11\times10^{12}{\rm{GeV}}$.
When these values are inserted into Eq. \eqref{max}, the resulting mass of the axion star is roughly $5.24\times10^{-10}M_\odot$. This mass is significantly lower than that of Planet 9, indicating that QCD axion stars cannot account for the presence of Planet 9.

Given that the mass of Planet 9 is at least $5M_\oplus$, this implies that the associated axion mass would be approximately $3.19 \times 10^{-10}$ eV, while the corresponding decay constant would be approximately $3.98 \times 10^{13}$ GeV if axions are to constitute the primary component of dark matter, as shown in Fig.~\ref{fig:constraints}. By substituting the axion mass and decay constant into Eq. \eqref{min}, we obtain a radius of $6.14\times10^4 \rm{km}$ for Planet 9, which is only slightly larger than that of Saturn. However, it's worth noting that the frequency of decay photons resulting from this scenario is too minimal to be detected by current radio telescopes.

\begin{figure}
\begin{center}
\includegraphics[width=0.5\textwidth]{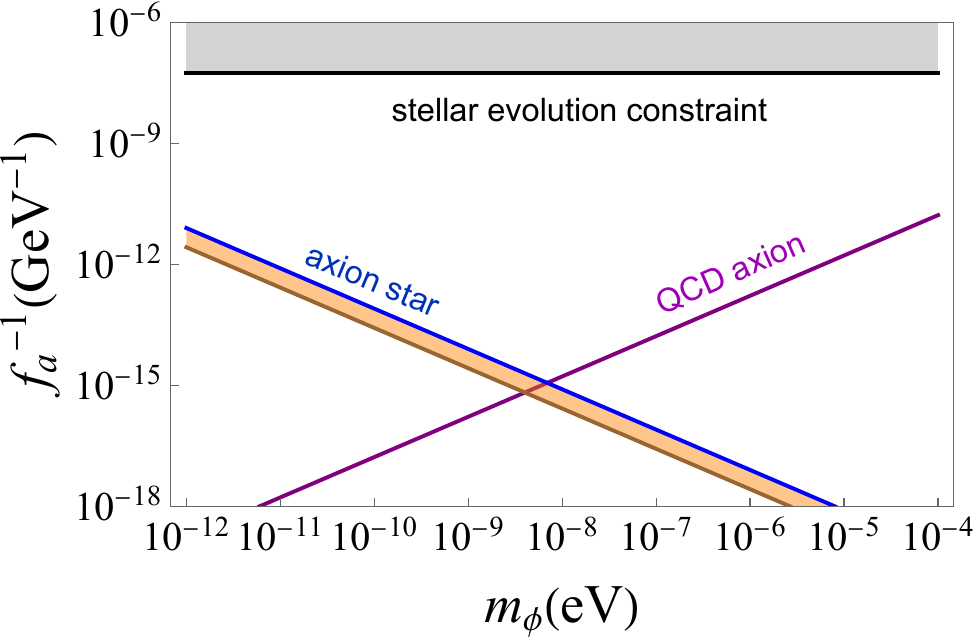}
\caption{Constraints on the parameter space of axions without considering the limitations of Eq. \eqref{abundance} due to some uncertainties, see Ref. \cite{Sugiyama:2021xqg} and the references therein. It can be seen that the QCD axion star can form the planet 9, with QCD axion mass $\sim10^{-8}{\rm{eV}}$ and $f_a\sim 10^{15}{\rm{GeV}}$. Nonetheless, these axion stars would still remain undetectable by current radio
telescopes. Furthermore, it's important to note that the luminosity of the corresponding axion star would fall below the sensitivity threshold of the FAST or SKA for the range of axion masses that can be explored by radio telescopes.}
\label{fig:constraints2}
\end{center}
\end{figure}

The aforementioned discussion relies on the constraint provided by the cosmological abundance as expressed in Eq. \eqref{abundance}. However, it's important to acknowledge that there exist uncertainties associated with this cosmological abundance limit, as discussed in Ref. \cite{Sugiyama:2021xqg} and  the references therein. For the purposes of our subsequent analysis, we will omit consideration of this abundance limit. When the QCD axion possesses a mass of approximately $10^{-8}{\rm{eV}}$ and a decay constant around $10^{15}{\rm{GeV}}$, it becomes possible for QCD axion stars with radius $78\rm{km}$ to give rise to the characteristics of Planet 9, as shown in Fig.~\ref{fig:constraints2}. Nonetheless, these axion stars would still remain undetectable by current radio telescopes. As a result, we intend to perform further calculations on the luminosity and flux of axion stars within a mass range that could potentially fall within the scope of observation by radio telescopes.
To determine the luminosity of Planet 9, we substitute the mass of the axion star, around $10^{-5}M_\odot$, and Eq. \eqref{max} into Eq. \eqref{luminosity}:
\bea
L_{\phi}=7.13\times10^{-42}\left({m_{\phi}\over {10^{-6}\rm eV}}\right)^3L_{\odot}.
\eea
Hence, the luminosity flux of the axion star as perceived from Earth is given by the formula:
\be
F_\phi={L_\phi\over 4\pi r_9^2},
\ee
where $r_9$ signifies the distance between the axion star and Earth. The resulting observable flux is then computed as follows:
\bea \label{flux}
F_{\phi}=9.70\times10^{-49}r^{-2}\left({m_{\phi}\over {10^{-6}\rm eV}}\right)^3{\rm (W/cm^2)}.
\eea
Here, the parameter $r$ assumes a value within the range of approximately $0.3$ to $1$, reflecting the uncertainty pertaining to the distance separating Planet 9 and Earth.

Radio telescopes can search for the potentially observable signal, which would appear as a monochromatic radio line at a universal frequency
set by the unknown axion mass.
The spectral flux density $S_{\rm{min}}$ is a crucial measurement in radio astronomy that indicates the minimum detectable signal strength for a radio telescope when observing normal electromagnetic signals, given by the following equation:
\bea
S_{\rm{min}}&=&\frac{\rm{SEFD}}{\eta_s \sqrt{n_{\rm{pol}}\mathcal{B}t_{\rm{obs}}}}
\eea
where SEFD refers to the equivalent flux density of a radio telescope system,
$n_{\rm{pol}}$ denotes the number of polarizations, $\eta_s$ represents the system
efficiency, $t_{\rm{obs}}$ is the observation time, and $\mathcal{B}$ is the bandwidth.

Five-hundred-meter Aperture Spherical radio Telescope (FAST) is currently the world's largest filled-aperture radio telescope. The frequency range of FAST is 0.1-3GHz, bandwidth is 800MHz, and SEFD is $2.2\rm{Jy}$ \cite{SKA:2015}. The spectral flux density $S_{\rm{min}}$ of FAST is
\bea
S_{\rm{min}}=3.60\times10^{-33}\rm{W/m^2/Hz},
\eea
where the number of polarizations $n_{\rm{pol}}=36$ for FAST. And we have assumed the system efficiency $\eta_s\simeq0.6$, observation time $t_{\rm{obs}}=1 \rm{hour}.$ In the dilute axion star, it's assumed that the axions remain nonrelativistic, thereby leading to a small velocity dispersion $\delta v\ll 1$,  which in turn results in a minimal spectral line spread. By employing the natural width of $10^{-6}f$ for the spectral line associated with the decay of the axion star, the expression for the minimum detectable flux can be derived as follows:
\bea
F_{\rm{min}}&=&3.60\times10^{-43}\left(\frac{f}{\rm{Hz}}\right)\rm{W/cm^2}\nonumber\\
&=&4.36\times10^{-35}\left(\frac{m_\phi}{10^{-6}\rm{eV}}\right)\rm{W/cm^2},
\eea
which gives the sensitivity of FAST. Compared with Eq. \eqref{flux}, it can be found that $F_{\phi}$ is much smaller than $F_{\rm{min}}$ and therefore the decay signal cannot be detected by FAST.

Square Kilometer Array (SKA) includes SKA1-Low and SKA1-Mid telescopes.  The SKA1-low covers 0.05-0.35 GHz, with $\mathcal{B}=300 \rm{MHz}$ and $\rm{SEFD}=4.9\rm{Jy}$. The SKA1-Mid covers 0.35-14GHz, with $\mathcal{B}=770 \rm{MHz}$ and $\rm{SEFD}=1.8\rm{Jy}$.  Relevant parameters of SKA can be found in Ref. \cite{SKA:2015}. The spectral flux density $S_{\rm{min}}$ of SKA1-Low and SKA1-Mid is $S_{\rm{min}}=1.31\times10^{-32}\rm{W/m^2/Hz}$ and $S_{\rm{min}}=3.00\times10^{-33}\rm{W/m^2/Hz}$ respectively, where we adopted the same integration time and system efficiency as FAST. The sensitivity of SKA1-Mid is similar to FAST and therefore the decay signal of axion star cannot be detected by SKA.

\section{Conclusions}
 QCD axion or ALP is a viable candidate for dark matter. A collection of axions can condense into bound BEC called axion star. It is possible that a significant fraction of the axion dark matter is in the form of axion stars. We find that the probability of capturing an axion star by the solar system is the same order of magnitude as the probability of capturing an FFP, and even higher for the case of axion star, with axion star mass $5M_\oplus\approx1.5\times10^{-5}M_\odot$ and $\Omega_{\rm{AS}}/\Omega_{\rm{DM}}\simeq 1/10$. Therefore, the anomalous orbits of TNOs can be explained by a QCD axion star or ALP star. However, unfortunately, the frequency corresponding to the decay of the axion star is either not within the frequency range of the radio telescopes such as FAST or SKA, or the decay signal is too weak to be detected. Therefore, if Planet 9 is composed by a QCD axion star or ALP star, it may be difficult to distinguish it from an isolated PBH.

 Moreover, if the axion star, such as Planet 9, exceeds the critical mass through accretion of axions from the background \cite{Chen:2020cef,Chan:2022bkz,Dmitriev:2023ipv}, it will collapse under the self-interaction of attraction, radiating radio waves through parametric resonance \cite{Hertzberg:2018zte,Levkov:2020txo} or generating relativistic axions \cite{Levkov:2016rkk}, which may be detected by upcoming experiments \cite{Arakawa:2023gyq}. We will study other observational effects of axion stars, including gravitational waves, in future work.

\section{Acknowledgments}

We would like to thank Xusheng Liu and Shaowei Jia for useful discussions. This work was supported by National Natural Science Foundation of China under Grant No. 11947031, and in part by East China University of Technology Research Foundation for Advanced Talents under Grant No. DHBK2019206.

\end{document}